\documentclass[prd,twocolumn,floatfix]{revtex4}
\usepackage[dvips]{graphics}
\usepackage{amsmath,amsfonts}
\usepackage{psfig}

\begin{document}

\newcommand{\bra}{\langle}
\newcommand{\ket}{\rangle}
\newcommand{\beq}{\begin{equation}}
\newcommand{\eeq}{\end{equation}}
\newcommand{\be}{\begin{equation}}
\newcommand{\ee}{\end{equation}}
\newcommand{\bea}{\begin{eqnarray}}
\newcommand{\eea}{\end{eqnarray}}
\def\bea{\begin{eqnarray}}
\def\eea{\end{eqnarray}}
\def\tr{{\rm tr}\,}
\def\erf{{\rm erf}\,}
\def\sgn{{\rm sgn}\,}
\def\href#1#2{#2}

\preprint{UW/PT-02/??}

\title{Experimental Tests of the Holographic Entropy Bound}

\author{Andreas Karch} \email{karch@phys.washington.edu}

\affiliation{Department of Physics,
University of Washington,
Seattle, WA 98195, USA}

\begin{abstract}
Kovtun, Son and Starinets proposed a bound on the viscosity of any
fluid in terms of its entropy density. The bound is saturated by 
maximally supersymmetric theories at strong coupling, 
but can also easily be challenged
experimentally to within a factor of 10 already today. 
We argue that this bound follows
directly from the generalized covariant entropy bound, bringing
holography within the reach of experimental investigation.

\end{abstract}
\today
\maketitle

\section{Introduction}

Kovtun, Son and Starinets recently proposed \cite{son} that in
nature the value\footnote{
From now on we set
$\hbar=k_B=1$.} \be \frac{\eta}{s} = \frac{\hbar}{4
\pi k_B} \approx 6.08 \, \times \, 10^{-13} K \cdot s, \ee where
$\eta$ is the shear viscosity of a fluid and $s$ its entropy
density, plays a fundamental role. They showed that idealized
systems, such as strongly coupled ${\cal N}=4$ supersymmetric
Yang-Mills theory and similar theories in 3 and 6 dimensions
precisely realize this value, while they conjectured that in
general it provides a {\it lower} bound. One
can easily envision that this hypothesis could be
tested experimentally in the near future. While for example water
under normal conditions (298.15 K, atmospheric pressure) has a
value of $\frac{\eta}{s} = 2.3 \times 10^{-10} Ks$, for superfluid
helium $^4$He one can obtain values down to $\frac{\eta}{s} = 6.20
\times 10^{-12} Ks$, only a factor of 10 away from the bound
\cite{son,handbook}.

What we will argue in this paper is that 
the viscosity bound
can be thought of as a consequence of the holographic entropy bound, or more
precisely from the generalized covariant entropy bound (GCEB) of
\cite{marolf}. Holographic entropy bounds are believed to be a
generic and fundamental property of any theory of quantum gravity,
(see \cite{thooft,susskind} or \cite{boussor} for a review).
Roughly speaking, the statement is that the number of degrees of
freedom in the universe don't scale as the volume of the universe,
but only as its surface area. A precise version of this
hypothesis, the Covariant Entropy Bound (CEB), has been formulated
by Bousso \cite{bousso}. The GCEB is an even stronger version of
this bound.

So far all these bounds have been motivated by theoretical
necessity. While many accept the CEB, the status of the GCEB
even among believers 
is somewhat less clear. Experimental tests of the entropy bounds
are in principle possible\footnote{For example it was argued
in \cite{hogan} that in principle the
CEB implies a discreteness in the temperature variations of the
cosmic microwave background.}. But one runs into the usual problem that
current experiments are orders of magnitude away from testing
quantum gravity. Showing that the GCEB implies the viscosity
bound of \cite{son} would bring testing holography into experimental
reach. We make the first steps into this direction.

\section{Hydrodynamics and Viscosity}

To understand the statement of the bound, we have to briefly
introduce the basic notions of hydrodynamics and how it
relates to strongly coupled systems with, following
the nice discussion in \cite{lgy}. Hydrodynamics should be
viewed as an effective theory, describing small fluctuations around
equilibrium of a thermal system
on length and time scales which are large compared to any microscopic scale in 
the system. The relevant hydrodynamic 
degrees of freedom are the conserved charge densities,
in the simplest case just for the charges derived from the energy-momentum
tensor, $\epsilon \equiv T^{00}$ and $\pi^i \equiv T^{i0}$. 
One equation of motion is just the current
conservation equation. In addition, one writes down
the so called constitutive relations, expressing the
fluxes of the conserved quantities (the spatial parts of the
conserved currents) in terms of the hydrodynamic degrees of freedom.
Without any further input, one has to allow here the most general
set of terms consistent with symmetries. Since hydrodynamics
is supposed to be an effective theory of long distances and small fluctuations,
one can systematically perform a double expansion in powers of the
fields and numbers of derivatives. To linear order
in the fluctuations and to first order in derivatives,
the constitutive relation for the spatial components of
the stress-energy tensor in $D$ spacetime dimensions reads
\bea
\label{const}
T^{ij}& =& \delta^{ij} \left ( P + v_s^2 \delta \epsilon \right )
- \gamma_{\zeta} \delta^{ij} {\vec{ \nabla} \cdot \vec{\pi}} \nonumber \\ 
&-& \gamma_{\eta} \left (
\nabla^i \pi^j + \nabla^j \pi^i - \frac{2}{D-1} { \vec{\nabla}
\cdot \vec{\pi}} \right ) .
\eea
$P$ is the equilibrium pressure. 
$\delta \epsilon$ denotes the fluctuation in $T^{00}$ around the
equilibrium value. For an equilibrium system in its rest frame, the
momentum densities are zero, so that the $\pi$'s are already the
first order fluctuations.
The three transport coefficients, $v_s^2$, $\gamma_{\eta}$ and $\gamma_{\zeta}$
depend on the microscopic details of the theory. The terms
linear in the momentum densities have been split into two independent
tensor structures.
The speed of sound $v_s$
and the bulk viscosity coefficient
 $\gamma_{\zeta}$ govern the propagation and diffusion of
longitudinal momentum fluctuations which form a coupled system with $\delta 
\epsilon$, 
while the shear viscosity
coefficient $\gamma_{\eta}$ governs the diffusion of transverse fluctuations
with $\delta \epsilon = \vec{\nabla} \cdot \vec{\pi} =0$,
which to this order decouple from the sound waves.
Conventionally one refers to the corresponding quantities
$\eta = (\epsilon + P) \gamma_{\eta}$ and $\zeta =
(\epsilon +P) \gamma_{\zeta}$ as the bulk and shear
viscosity respectively. They would appear directly in the constitutive
relation if we were to linearize in velocities instead of momenta.
In a conformal theory like maximally supersymmetric Yang Mills theory,
the stress-energy tensor has to be traceless, implying that
$\zeta=0$ and $v_s=\frac{1}{\sqrt{D-1}}$. The beauty
of the hydrodynamic description is that it applies to any system
which acts like a fluid at long distances, even strongly coupled
ones. For weak coupling $\eta$ can be calculated perturbatively.
For strongly coupled theories one has to find alternative
techniques. As shown in \cite{son}, at weak coupling $\frac{\eta}{s} >>1$,
so that the bound only gets tested by strongly coupled systems.

\section{General Strategy for Derivation of the Bound}

In order to understand how viscosity could appear in holographic
entropy bounds, let us briefly recapitulate how gravity enforces
such bounds. If we had a fluid in flat space at rest with a finite
entropy density, one could obviously violate the bound. The bound states
that the entropy passing the lightsheet, which is constructed by emitting
lightrays from the boundary of a given volume element inward, should be less
than the area divided by $4 G \hbar$. For a volume element of the fluid
in flat space,
the whole fluid enclosed in a given area passes the
lightsheet, so that the entropy should scale with the volume by
the assumption that we have a finite entropy density. The way
gravity avoids this conflict is that it makes it inconsistent to
have such a fluid in {\it flat} space. The energy stored in the
fluid will curve space-time sufficiently to rescue the holographic
bound.

\begin{figure}
 \centerline{\psfig{figure=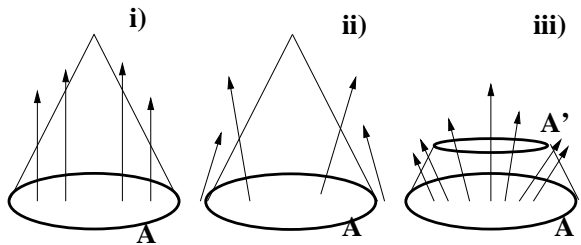,width=3.0in}}
 \caption{i) illustrates the CEB: all matter contained within
the area $A$ has to pass the lightsheet, and this is still true
even if there are velocities and velocity gradients as in ii).
However in the case of the GCEB, where we are only
interested in entropy passing through the lightsheet from $A$ to
$A'$, matter inside both $A$ and $A'$, that would not have
been counted at rest can move into the lightsheet as
depicted in iii). It is
only in this case that velocity gradients give new challenges
to the bound.
}
 \label{strategy}
  \end{figure}
Viscosity becomes important, once the fluid has non-trivial
velocity profiles. For the CEB, it does not seem to matter much
whether the fluid is in motion: focus on a certain volume element
of fluid. No matter what the velocities are, the whole fluid will
eventually pass the lightsheet. On the other hand, no particles
that where originally outside the volume can move inside fast enough to 
be counted by 
the lightsheet, since
they are slower than light. The situation however is quite
different for the GCEB. Here we are dealing with lightsheets that
terminate, and the entropy is bounded by the area the lightsheet emanates
from {\it minus} the area in which we chose to terminate it. 
Parts of the fluid that start out inside but sufficiently close to
the spatial location at which we chose to terminate the lightsheet
will be able to move outside the region before the light arrives
there and avoid to be counted. Similar, fluid that starts 
outside the region we want to sample can move inside and might
lead to a violation of the bound. This situation
is sketched in Fig.\ref{strategy}. As before, it is gravity that
has to censor any possible violation. What we need is that
velocity profiles lead to a curved spacetime, in which the
lightsheet gets sufficiently modified. This is precisely where
viscosity enters: from the constitutive relation we see that
viscosity tell us how much stress $T_{ij}$ we have in a fluid once
we turned on the velocity profile. And by Einstein's equations
this stress $T_{ij}$ will turn on spacetime curvature $R_{ij}$. In
order to not violate the bound, the backreaction has to be
sufficiently large, that is for a fluid with given entropy
density, the viscosity is not allowed to become too small. In the
next section we will show how this works quantitatively.

\section{Derivation of the Viscosity Bound}
\subsection{Review of Bousso's Derivation of Bekenstein's Bound}
In \cite{boussob} Bekenstein's bound on the entropy of a matter system
was derived from the GCEB. The setup used there can easily be generalized
to apply to the case of a fluid in motion.
\begin{figure}
 \centerline{\psfig{figure=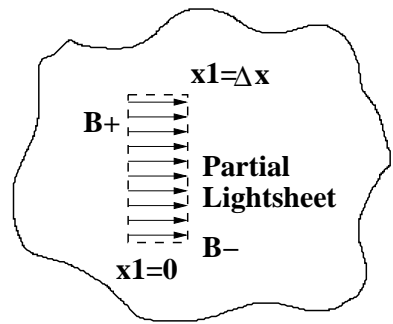,width=2.0in}}
 \caption{
A large droplet of fluid, out of which a small volume element gets
sampled by a lightsheet starting at $B_+$ and terminated at $B_-$.}
 \label{fluid}
  \end{figure}
Consider a large region of spacetime occupied by a fluid. We are interested
in the entropy contained in a volume element bounded by surfaces $B_+$ 
and $B_-$. A lightsheet (a null-hypersurface generated by non-expanding
light-rays) enters the volume element at $B_+$ and exits at $B_-$. 
The GCEB states
that
\be S \leq \frac{A(B_+) - A(B_-)}{4 G } \label{gceb} \ee
where $S$ is the entropy passing through the lightsheet, and $A(B_i)$ denotes
the area of the corresponding surface.
For
simplicity we will work with rectangular surfaces as depicted in
Fig.\ref{fluid}, but the result is independent of the shape, since the areas
involved do not change if we deform $B_{\pm}$ along the lightsheet.
As in \cite{boussob} we chose the lightsheet to be given by the set
of parallel light rays obeying
\be x^0 = x^1; \, \, \, (x^2, \ldots, x^{D-1}) \mbox{ arbitrary constants.}
\ee 
These are null geodesics in flat space. For the same reason that
the area is invariant under deformations along the lightsheet we
actually have $A(B_+)=A(B_-)$ and the entropy seems to be zero. In
order to get a non-vanishing area difference we have to take into account the
backreaction of the matter system. Due to the stress energy $T_{\mu \nu}$
of the fluid, non-trivial curvature $R_{\mu \nu}$ will be turned on, and
the above hypersurface will neither be null nor geodesic. However, for weakly
gravitating systems, there will be a close by null geodesic lightsheets.
\cite{boussob} uses two such close by lightsheets, one that starts in B$_+$ 
but misses $B_-$, and one that ends in $B_-$, but does not start at $B_+$.
Plugging those into eq.\ref{gceb}, one obtains the following bound ($\hbar=1$):
\be S \leq \pi \, \Delta x \, \int dx_2 \ldots dx_D 
\int_{0}^{\Delta x} dx_1 T_{\mu \nu} k^{\mu} k^{\nu}, \ee
where $k^{\mu \nu}=(1,1,0,0)$ is a tangent null vector to the lightsheet.
For the volume element of our fluid the rhs evaluates to 
\be S_{eq.} = \pi V (\epsilon + P) \Delta x.\ee
What will be important in this expression later is that it is proportional
to $\Delta x$, so that the equilibrium configuration will lead to
zero entropy in infinitesimally thin spatial slices. In \cite{boussob}
it was further observed that if one were to chose a lightsheet that includes
the whole region of spacetime occupied by the fluid,
$\int dx_2 \ldots dx_D 
\int_{0}^{\Delta x} dx_1 T_{\mu \nu} k^{\mu} k^{\nu} $ just evaluates
the total ADM mass of the fluid in its rest frame and one can recover
the conventional form of Bekenstein's bound \cite{bek}, $S\leq \pi M \Delta x$.

\subsection{Letting Things Flow}
In order for viscosity to enter the game we have to allow for
motion of the fluid. Consider a fluid as described above, but turn
on a perturbation away from equilibrium which at $t=0$ has the
following momentum density profile: \be \pi_1(\vec{x},t=0) = -
\sgn(x_1-\Delta x/2) \, \cdot \,  (\epsilon + P) \, \cdot \, v_0 \ee where $\epsilon$ and
$P$ are the equilibrium values of the energy density and the
pressure. \be \label{sol} v_0 \leq 1 \ee is the velocity at which
the fluid is moving with respect to the restframe of the
equilibrium system (which has to be less than the speed of light).
Strictly speaking, as $v_0$ approaches 1 (or if we go out to
large values of $x_2$), we would have to use
relativistic hydrodynamics instead and, in addition, are no
longer justified to neglect the higher order perturbations in the
constitutive relation. So our derivation of the bound only applies
to non-relativistic systems, even though we expect the bound
to be also true in the relativistic case.

We have set up the system in such a way that the fluid at negative
$x_1$ is moving to the right (in vane trying to escape the
lightsheet), while the fluid at positive $x_1$ is moving to the
left (rushing into the lightsheet). In addition we turn on \be
\pi_2 (\vec{x},t=0) = 2 \delta(x_1 - \Delta x/2) \, \cdot \, x_2
\, \cdot \, (\epsilon +
P) v_0 \ee so that \be \vec{\nabla} \cdot \vec{\pi} =
\partial_1 \pi_1 +
\partial_2 \pi_2 =0 \ee and we are dealing with a purely
transverse fluctuation. The linearized
hydrodynamics can be solved in this case and the full 
time dependent solution is
obtained by having the stepfunction diffuse with diffusion
constant $\gamma_{\eta}$, \beq
\pi_1(\vec{x},t) \sim \erf(\frac{x - \Delta x/2}{\sqrt{4 \gamma_\eta t}}) .\ee

In order to calculate the maximal entropy density allowed within
the lightsheet bounded by $0<x_1<\Delta x$, we follow the same
logic as in the derivation of Bekenstein's bound and find once
more \bea s &\leq& \frac{\pi}{V} \int dx_2 \ldots dx_D \Delta x
\int_{0}^{\Delta x} dx_1 T_{ab} k^a k^b \\ &=\pi &
\int_0^{\Delta_x} dx_1 (T_{00} + 2 T_{01} + T_{11}). \eea The
equilibrium values on the right hand side as before yield $s_{eq.}
=\pi \Delta x \, (\epsilon + P)$. In the
same way, the contribution from $T_{01}$
gives an integral over the momentum density $\pi$. The way we set
up the profile and the lightsheet geometry, the net momentum in our
lightsheet is actually zero. But even in a more general configuration,
the $T_{01}$ term will always give an integral over the local momentum
density and vanish as we take $\Delta x$ to zero, which as we will soon
see is all we need. Last but not least let's calculate what we
get from the linear perturbation $\delta T_{11}$ 
in the $T_{11}$ piece. From the
constitutive relation eq.\ref{const} we see that \be \delta T_{11} = - 2
\gamma_{\eta}
\partial_1 \pi_1(\vec{x},t) \ee and on the right hand side
we get a contribution \be s \leq s_{eq.} + \pi 4 v_0 \eta
\int_0^{\Delta x} dx_1 \delta(x_1 - \Delta x/2) .\ee 
We have neglected the fact that the delta function
diffuses while being sampled by the lightsheet. Since total momentum
is conserved under diffusion, the integral over
the full delta function gives an upper bound on the integral
of the time dependent solution\footnote{
The diffusion process will also generate entropy which
can be neglected as long as $\Delta t=
\Delta x/c$ is much smaller than the diffusion timescale.
See also the next footnote.} 

The important thing to note is
that the viscosity term enters via an integral of a
delta-function. We only get a surface contribution
from the boundary of the lightsheet, the result is independent of $\Delta x$. Since the
bound has to be satisfied for arbitrary lightsheets, it has to be
true for lightsheets with infinitesimal $\Delta x$. In this case the bulk
contributions, that is
both the equilibrium contribution as well as the integral over the
momentum density $T_{01}$, drop out\footnote{
Unfortunately for a realistic matter system, for which the
fluid form of the stress tensor is only a long distance approximation,
this limit seems to force us to take $\Delta x$ so small that hydrodynamics
is not necessarily a good description. One certainly can still setup a flow
with the initial conditions we use, but to what extend it is the viscosity
alone that governs the stress is up to debate.}.
Using further $v_0 \leq 1$ we
obtain the viscosity bound of \cite{son},
$s \leq 4 \pi \eta$ .

\section{Interpretation and Future Direction}

Should one really interpret the viscosity bound as being some
subtle imprint of quantum gravity on properties of macroscopic
systems? For a somewhat more conservative point of view to read
our results, recall the connection between Bekenstein's
bound and the GCEB: the hypothetical holographic bound, involving
both Newton's constant $G$ and Planck's constant $\hbar$ gets
combined with formulas of classical general relativity involving
only $G$ yielding a bound that only involves $\hbar$ and turns out
to basically correspond to Heisenberg's uncertainty principle.
While one could read this as ``Holography implies the uncertainty
principle'' a more conservative way to read this result is to say:
once we chose to deal with quantum gravity, we obviously also have
to treat the matter fields quantum mechanical, and the usual rules
(including the uncertainty principle) will apply. Latter can be
understood without ever appealing to quantum gravity.

In the same spirit we think one should read the viscosity bound.
The GCEB can be used to derive a property of macroscopic systems,
which presumably can also be derived by other methods. The reason
it had not been obtained earlier is that it is a strong coupling
phenomenon. As shown in \cite{son}, in a weakly coupled field
theory, $\frac{\eta}{s} >> 1 $.
The supersymmetric models saturating the bound are
formally at infinite coupling, and the only reason one was able
to compute $\frac{\eta}{s}$ is that they have a weakly coupled
gravity dual. Still, the viscosity bound can be challenged
experimentally, and any experimental test 
automatically becomes a test of the GCEB. For the
future it might be very interesting to study in the same spirit
non-vanishing chemical potentials or charged fluids in background fields. 
Most
likely one can derive bounds similar to the viscosity bound from
the GCEB that govern the behavior of other transport coefficients,
giving even more opportunities to probe the GCEB in the
laboratory.

\vskip15pt

\noindent{\large\bf Acknowledgments:} We would like to thank
Raphael Bousso, Josh Erlich and Pavlo Kovtun for useful discussions. 
This work was
partially supported by the DOE under contract DE-FGO3-96-ER40956.

\bibliography{bound}
\bibliographystyle{apsrev}
\end{document}